\documentclass[epj,twocolumn]{webofc}
\usepackage[varg]{txfonts}   
\def\sch{Schwarzschild}
\woctitle{The innermmost regions of relativistic jets and their magnetic fields  }
\begin{document}
\title{Jets, black holes and disks in blazars}
%
%

\author{Gabriele Ghisellini  
\inst{1}\fnsep \thanks
{\email{gabriele.ghisellini@brera.inaf.it}} 
}

\institute{ INAF -- Osservatorio Astronomico di Brera
          }

\abstract{
The {\it Fermi} and {\it Swift} satellites, together with ground based Cherenkov telescopes, 
has greatly improved our knowledge of blazars, namely Flat Spectrum Radio
Quasars and BL Lac objects, since all but the most powerful
emit most of their electro--magnetic output at $\gamma$--ray energies,
while the very powerful blazars emit mostly in the
hard X--ray region of the spectrum. 
Often they show coordinated variability at different frequencies,
suggesting that in these cases the same population of electrons is at work,
in a single zone of the jet.
The location of this region along the jet is a matter of debate.
The jet power correlates with the mass accretion rate,
with jets existing at all values of disk luminosities, measured in 
Eddington units, sampled so far.
The most powerful blazars show clear evidence of the emission from their
disks, and this has revived methods of finding the black hole mass
and accretion rate by modelling a disk spectrum to the data.
Being so luminous, blazars can be detected also at very high redshift,
and therefore are a useful tool to explore the far universe.
One interesting line of research concerns how heavy are their black holes at high redshifts. 
If we associate  the presence of a relativistic jets with a fastly spinning
black hole, then we naively expect that the accretion efficiency is 
larger than for non--spinning holes.
As a consequence, the black hole mass in jetted systems should grow at a slower rate.
In turn, this would imply that, at high redshifts, the heaviest black holes 
should be in radio--quiet quasars.
We instead have evidences of the opposite, challenging our simple ideas of 
how a black hole grows. 
}
\maketitle
\section{Introduction}
\label{intro}
Blazars are extragalactic objects with a relativistic jet pointing at us.
They often emit most of their electromagnetic output in the $\gamma$--ray band,
and they are the most numerous class of extragalactic objects detected 
by the Large Area Telescope (LAT) onboard the {\it Fermi} satellite \cite{nolan12} 
and by on ground Cherenkov telescopes (see \cite{hinton09} for a review).
These relatively recent $\gamma$--ray data improved
the knowledge of the basic properties of blazars, even if,
as usually happens, new open issues appeared.
Perhaps one of the most puzzling is where, in the jet, most of the
luminosity is produced.
As discussed below we have conflicting evidences, and
suggestions ranging from 0.1 to 10 parsecs.

Another long debated issue is the real existence of a {\it blazar sequence},
in which low power blazars are characterized by a Spectral Energy Distribution (SED)
with two broad hump peaking respectively in the UV--soft X--ray and in the GeV--TeV
bands, while high power blazars peak at smaller frequencies (sub--mm and $\sim$MeV;
\cite{fossati98} \cite{donato01} 
but see e.g. \cite{padovani07} 
and \cite{giommi10} 
for an alternative view).

Blazars can be divided into Flat Spectrum Radio Quasars (FSRQs) and BL Lac objects 
according to the equivalent width (EW) of their emission lines,
greater or smaller than 5 \AA\ \cite{urry95}, respectively.    
This phenomenological division, although useful in practice,
leads to several misclassifications, since a particularly intense
(and possibly beamed) continuum can hide the broad lines.
An intrinsically FSRQ is then misclassified as a BL Lac.
On the other hand, in very low states, the EW of a 
``genuine" BL Lac can occasionally exceed 5 \AA, as
occurred to BL Lac itself \cite{vermeulen95}.
We (\cite{gg11}, \cite{sbarrato12a}) 
have proposed a more physical classification, based on the
broad line luminosity measured in Eddington units $L_{\rm BLR}/L_{\rm Edd}$.

The radiation produced by relativistic jets
is strongly boosted by beaming, making blazars very bright
even at high redshifts.
Therefore they can be used as cosmological tools to explore the far Universe.
The most powerful blazars are the ``reddest", with a synchrotron peak
in the sub--mm band, and leave the accretion disk emission unhidden.
For them, we can easily and confidently find the black hole mass and 
accretion rate.
Therefore a very promising line of research concerns the hunt of heavy black holes
at redshifts greater than 4.
The physical motivation is twofold: on one hand discovering a blazar 
(whose jet is aligned with our line of sight) with a heavy black hole
implies the existence of hundreds of similar black holes with jets pointing
elsewhere.
Discovering a single blazar at $z\sim 6$ with a black hole mass $M>10^9 M_\odot$
implies the existence of 450$\cdot(\Gamma/15)^2$ black holes with the same mass.
This outnumbers the known heavy black holes in radio--quiet quasars at the same redshift
(there are $\sim$30 radio--quiet quasars with a spectroscopic redshift
$z >6$ listed in NED, {\tt http://ned.ipac.caltech.edu}).

The second point concerns the growth of {\it spinning} black holes.
It is likely, in fact, that relativistic jets are associated with fast spinning holes. 
In these systems the accretion disk  is very radiatively efficient (disk luminosity
$L_{\rm d}=\eta \dot M c^2$, with $\eta\ge 0.1$ up to $\eta=0.3$; \cite{thorne74}).
For this accreting Kerr holes, the Eddington luminosity is reached with a 
a smaller $\dot M$ than for a \sch\ hole.
As a consequence an Eddington limited black hole grows at a slower rate, and 
cannot reach $10^9 M_\odot$ at $z>4$. 
{\it But we do see jetted sources with heavy black holes at $z>4$.}

\begin{figure}
\vskip -0.6 cm
\hskip -0.4 cm
\includegraphics[width=9cm,clip]{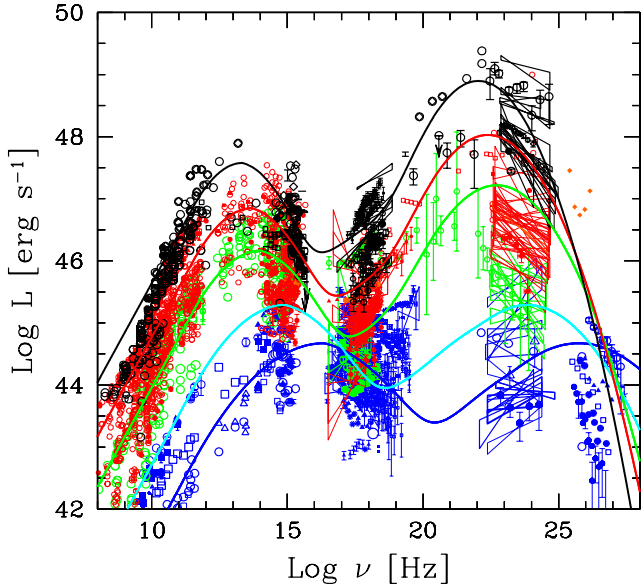}
\vskip -0.5 cm
\caption{
All blazar SED detected in the first three months of operations
divided according to their luminosity in the LAT
band (0.1--100 GeV): $\log L_\gamma<45.5$ (blue); $45.5\log L_\gamma<46.5$ (green); 
$46.5<\log L_\gamma<47.5$ (red); 
$\log L_\gamma>48$ (black). 
Solid lines are the phenomenological models in \cite{fossati98}, that divided 
the blazars according to their {\it radio} luminosity, assumed to be a good tracer of
the bolometric one. Luminosities in erg s$^{-1}$.
}
\label{gfos}       
\end{figure}

\section{The status of the blazar sequence}

Soon after the first detection of blazars in the $\gamma$--ray band by the EGRET instrument,
it was clear that the SEDs of blazars were characterized by two broad humps, interpreted
as the synchrotron and the inverse Compton emission.
Quite often the high energy hump was dominant.
Fossati et al. (1998, \cite{fossati98}, see also the update of \cite{donato01})
considered a few complete (in radio and in X--rays) blazar samples,
for a total of about one hundred objects, and divided them in 5 different bins of
radio luminosity. 
By considering photometric data at different frequencies for each object,
and averaging the fluxes at selected frequencies for the objects belonging
to the same radio bin, they constructed 5 typical SEDs.
The obtained SED described a sequence with the following properties:
i) the radio luminosity was a good tracer of the bolometric one;
ii) by increasing the radio (hence the total) luminosity, the frequencies 
of the two peaks shifted to smaller values, and, at the same time,
iii) the high energy peak became more important.
It was possible to describe the results with continuous functions, constructed
on a phenomenological basis, shown by the solid lines in Fig. \ref{gfos}.

\begin{figure}
\vskip -0.6 cm
\hskip -0.5 cm
\includegraphics[width=9.cm,clip]{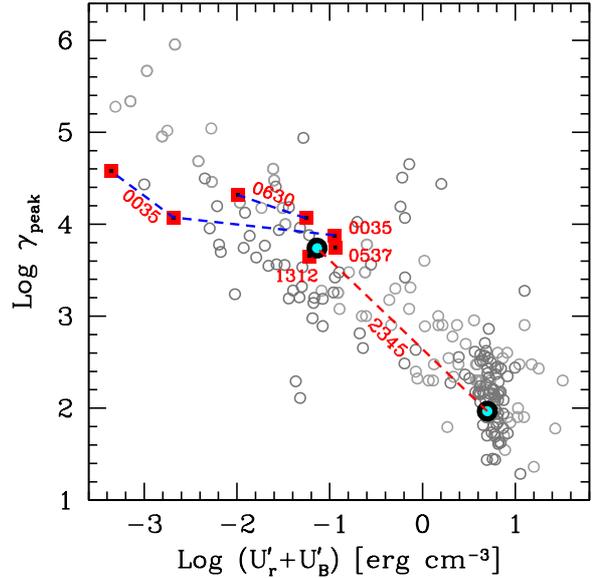}
\vskip -0.6 cm
\caption{
The (random) Lorentz factor $\gamma_{\rm peak}$ of the electron emitting at the 
peaks of the SED as a function of the total (magnetic plus radiative) energy density, 
as seen in the comoving frame of the jet.
Labels refer to specific objets (namely the four "blue" quasars discussed in \cite{gg12} 
and \cite{giommi12}. 
The dashed red line joins two states of the blazar J2345--1555 which became 
``blue" during a flare in 2013, see \cite{gg13a}. 
}
\label{gpeak}       
\end{figure}

\subsection{Interpretation}

The blazar sequence was interpreted by \cite{gg98} as
the result of the different amount of radiative cooling in different sources.
Low power sources are BL Lac objects, with weak or absent broad emission lines.
Being less powerful, their jet presumably carries a weaker magnetic field.
The main emission mechanisms are synchrotron and self Compton.
Since the cooling is limited, electrons can attain high energies,
and they preferentially produce high frequency radiation.
As a result, the produced SED is ``blue" (namely, the synchrotron
peak is at UV or soft X--ray frequencies, while the high energy peak
can reach the TeV band. 
These blazars are also called High frequency BL Lacs, 
or HBL, by \cite{padovani95}).
By increasing the total luminosity, we have objects with strong broad emission lines,
and presumably jets with stronger magnetic fields.
Cooling is more severe, and the electron energies are smaller.
The peak frequencies of the two humps shift to the ``red" (sub--mm
for the synchrotron, MeV for the Compton. 
These are called Low frequency BL Lacs, or LBL, by \cite{padovani95}).
At the same time, the electrons can scatter seed photons not only
produced internally in the jet (i.e., their own synchrotron photons),
but also the seeds coming externally (disk, broad line region, torus).
The enhanced abundance of seed photons makes the scattering process
more important, and the high energy bump is then dominant.
This can be demonstrated by applying simple, one--zone,
synchrotron + inverse Compton (leptonic) models to 
blazar samples (for the alternative hadronic models, and a nice comparison between
leptonic and hadronic jet models, see \cite{bottcher13}).
If one relates the energy of the electrons emitting at the peak of the SED
with the total (magnetic plus radiative) energy density as seen in the
comoving frame, one finds a strong correlation, shown in 
Fig. \ref{gpeak}: high electron energies are possible only if the energy densities
are small.
This confirms the idea that the blazar sequence is a by--product of the different
amount of radiative cooling.

In \cite{fossati98} we divided blazars according to their radio luminosity,
thought to be a good tracer of the bolometric one. 
Now, with the advent of the {\it Fermi} satellite, we can divide blazars according to their
$\gamma$--ray luminosity in the LAT band [0.1--100 GeV].
The result is shown in Fig. \ref{gfos}, and agrees remarkably well
with the phenomenological solid curves proposed 15 years ago.
There is a discrepancy between the phenomenological models and the $\gamma$--ray data
at intermediate luminosities, but this can easily be explained remembering that
the phenomenological models were built when only the brightest (and more luminous)
$\gamma$--ray blazars were known. 
The conclusion is that {\it Fermi} fully confirms the existence of the blazar sequence.

\begin{figure} 
\vskip -0.5 cm
\hskip -0.1 cm
\includegraphics[width=8.8cm,clip]{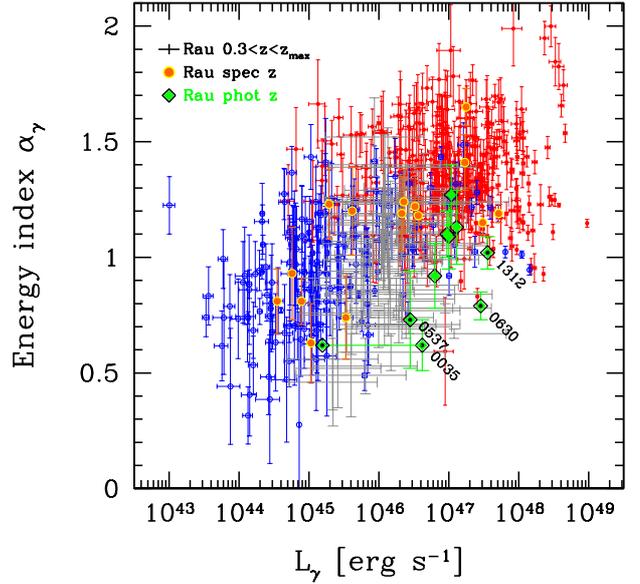}
\vskip -0.5 cm
\caption{
The energy spectral index $\alpha_\gamma$ as measured by {\it Fermi}
as a function of their [0.1--10 GeV] (rest frame) luminosity.
Red  points are FSRQs, blue circles
are BL Lacs.
The sources studied in \cite{rau12} are plotted with larger symbols
if the redshift is determined (spectroscopically or photometrically).
Sources with only un upper limit for the photometric redshifts
are plotted assuming a luminosity the upper limit as the maximum redshift,
and $z_{\rm min}=0.3$ as the minimum (this corresponds approximately
to the limit at which the host galaxy should be visible).  
}
\label{divide}       
\end{figure}

On the other hand, the number of {\it Fermi} sources with a featureless spectrum 
(thus of unknown redshift) is large.
These could be genuine BL Lacs without emission lines or FSRQs whose continuum is so beamed
to hide the broad emission lines. 
Many of these sources have a flat $\gamma$--ray spectrum 
($\alpha<1$ when $F_\nu \propto \nu^{-\alpha}$). 
As shown by Fig. \ref{divide} (empty blue circles), sources with flat $\gamma$--ray 
spectra are preferentially BL Lacs.
If they are at -- say -- $z=4$, then there is a population of BL Lac objects
that is very powerful, and nevertheless have a ``blue" spectrum.
This would contradict the blazar sequence.
Rau et al. (2012) \cite{rau12} succeeded to put limits on the redshifts of these objects,
by using the GROND instrument (giving 7 photometric data simultaneously,
from near infrared to optical) together
with UVOT (onboard {\it Swift}) giving 6 optical--UV photometric data.
If a source is at high--$z$, the intervening material imprints 
absorption features in the spectrum (Lyman--$\alpha$ forest and edge)
that can be used to estimate a photometric redshift.
Fig. \ref{divide} reports the results: the green diamond (with a black dot)
correspond to the sources with $z$ measured in this way.
At the same time, the {\it absence} of absorption features ensures that the object
is at a redshift smaller than a maximum value $z_{\rm max}$ (around 2).
This objects are shown with the grey bars, corresponding to values of the luminosity
calculated assuming a redshift between $z_{\rm max}$ and 0.3 (if smaller, one can
see the host galaxy).

In investigating the 4 sources with the flat $\gamma$--ray spectra and
the measured photometric redshift, \cite{gg12} 
concluded that 
they are FSRQs with relatively weak accretion disk luminosity and with a 
strongly boosted jet continuum. 
The energy $\gamma_{\rm peak}$ of the relevant electrons of these sources is
shown in Fig. \ref{gpeak}: they well agree with the
general correlation, but lie in the ``blue part" (high $\gamma_{\rm peak}$), 
where usually we find low power BL Lacs.
Another example is the blazar PMN J2345--1555: it is usually characterized by
a ``red" spectrum, but during a flaring episode, it became ``blue".
This was interpreted by \cite{gg13a} 
as a change in the location of the
dissipation region along the jet, where most of the flux is produced:
usually it is within the broad line region (BLR), where the presence of
broad line photons ensures a large radiation energy density and thus a small
$\gamma_{\rm peak}$ and a red spectrum.
Occasionally, the dissipation region may move beyond the BLR: there the
decreased radiation energy density let the electrons reach large 
$\gamma_{\rm peak}$ (Fig. \ref{gpeak}), making the spectrum blue.

\subsection{Dividing FSRQs and BL Lacs}

As mentioned in the introduction, the traditional division between 
FSRQs and BL Lacs is based on the EW of the broad emission lines.
Furthermore, the parent population (i.e. the misaligned counterparts)
of FSRQs is made by powerful FR II radio--galaxies, while the parents
of BL Lacs are low power FR I radiogalaxies. 
There are however several exceptions to this rule, and intermediate
objects exist.
Furthermore, according to \cite{scarpa97} 
there is a continuity between 
BL Lac and FSRQs in the MgII line luminosity with no clear separation of blazars 
in the two subclasses.

\begin{figure} 
\vskip -0.6 cm
\hskip -0.5 cm
\includegraphics[width=9.2cm,clip]{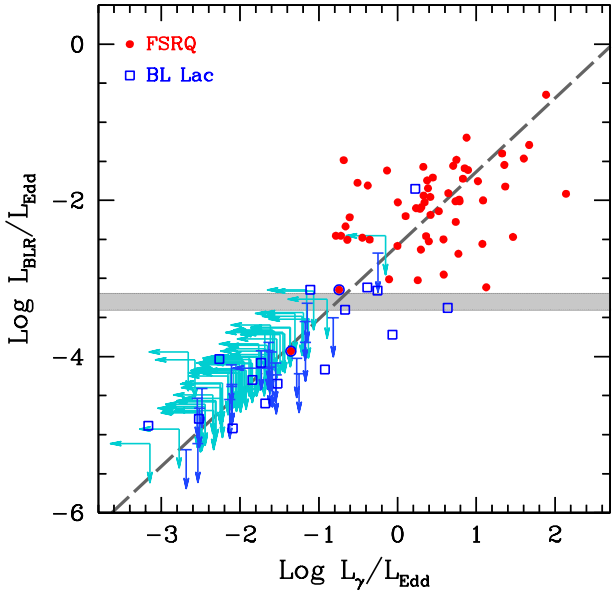}
\vskip -0.5 cm
\caption{
The luminosity of the broad line region as a function of the $\gamma$--ray luminosity,
when both are measured in Eddington units.
FSRQs (red circles) and BL Lacs (blue) are classified according to the classical definition
(EW of the broad lines larger of smaller of 5 \AA). 
The grey horizontal line marks the divide between FSRQs and BL Lacs, at 
$L_{\rm BLR}/L_{\rm Edd}=5\times 10^{-4}$.
Adapted from \cite{sbarrato12a}.
}
\label{lblrlg}       
\end{figure}

In \cite{gg11} and \cite{sbarrato12a}
we have studied the luminosity of the BLR ($L_{\rm BLR}$) as a function of the
$\gamma$--ray luminosity ($L_\gamma$). 
The first is a proxy of the disk luminosity, while the latter is a proxy
of the bolometric jet luminosity, in turn linked with the jet power.
We found that the two luminosities correlate, even when accounting for the common 
redshift dependence, and even considering that the $\gamma$--ray luminosity
can vary, in a single object, by more than two orders of magnitude.
Blazar classified as BL Lacs in the ``classical" way (EW$<$5\AA) have
small $L_{\rm BLR}$ and $L_\gamma$. 
Furthermore, since there are estimates 
of the black hole mass for a good fraction of the sample, we could
also see at which $L_{\rm BLR}/L_{\rm Edd}$ there is the divide between
BL Lacs and FSRQs. 
We found a value $L_{\rm BLR}/L_{\rm Edd}\sim 5\times 10^{-4}$, as shown 
in Fig. \ref{lblrlg}.
Since the disk luminosity $L_{\rm d}\sim$10--20$L_{\rm BLR}$, 
this is in good agreement with the value suggested earlier by \cite{gg09a} 
on the basis of the BL Lac/FSRQs division in the $\alpha_\gamma$--$L_\gamma$
plane, and by \cite{gg01} 
on the basis of the FR I/FR II division in the radio 
power -- galaxy bulge optical luminosity plane.

\subsection{Accretion and the blazar sequence}

For what discussed above, we can draw the conclusion that 
the jet power in blazars correlates with the accretion rate
of their accretion disks.
Since we also know the black hole masses of several blazars, 
we can also say that: 
i) the blazar' divide occurs at $L_{\rm d} \sim 10^{-2} L_{\rm Edd}$;
ii) jets are present {\it for all} $L_{\rm d}/L_{\rm Edd}$ we can measure, i.e. from 
     below $L_{\rm d}/L_{\rm Edd} \sim 10^{-4}$ to 1.
In addition, the jet power often exceeds the disk luminosity 
(\cite{celotti08}, \cite{gg10b}).
We can envisage two scenarios for explaining the FSRQs/BL Lacs divide:

\begin{enumerate}

\item At the critical luminosity $L_{\rm d}\sim 10^{-2} L_{\rm Edd}$, the disk accretion
changes nature, from being radiatively efficient to radiatively inefficient.
In other words the efficiency $\eta$ in $L_{\rm d} =\eta \dot M c^2$ is constant
($\eta\sim 0.1$) for $L_{\rm d}\ge 10^{-2} L_{\rm Edd}$, while it decreases
for lower values \cite{narayan95}, \cite{narayan97}. 
The disk becomes geometrically thick as the result of inefficient proton cooling, since
electrons and protons are decoupled, and protons remain hot.
The main radiation mechanism in these conditions are cyclo--synchrotron, 
bremsstrahlung and inverse Compton scattering \cite{mahadevan97}: 
the amount of UV photons to ionize the clouds of the BLR decreases dramatically.
Emission lines then becomes very weak or even absent.
The power of the jet, however, continues to be proportional to $\dot M$.
If in this regime $L_{\rm d}\propto \dot M^2$, as suggested by \cite{narayan97},
then $P_{\rm jet}/L_{\rm d}\propto \dot M$.

\item The divide between radiatively efficient and inefficient disks might occur
at values of $L_{\rm d}/L_{\rm Edd}$ smaller than $10^{-2}$.
Sharma et al. (2007) \cite{sharma07}, indeed, suggest a smooth transition around $10^{-4}$.
In this case there is not a sharp decrease of the UV ionizing radiation,
and weak broad emission lines continue to be produced.
The ionizing luminosity, however, correlates with the size of the BLR $R_{\rm BLR}$
(as $R_{\rm BLR} \propto L^{1/2}$), so in these sources $R_{\rm BLR}$ is small,
and it is very likely that the jet produces most of its non--thermal
luminosity {\it beyond} $R_{\rm BLR}$.
This could explain why broad lines are visible even in BL Lac itself or 
in Mkn 421 and Mkn 501 \cite{morganti92}, \cite{stickel93}.

\end{enumerate}

\section{One--zone model?}

In the early and mid `80s it was thought that the observed non--thermal luminosity
was produced all along an inhomogeneous jet (\cite{marscher80}, \cite{konigl81}, \cite{gg85},
see \cite{potter13} for a revival)  
with magnetic field and particle density that were power law functions of the 
distance from the black hole.
As a consequence, the flux at different frequencies was produced mainly 
at different jet locations, with different variability timescales.
The high energy hump of the SED was not yet discovered, and 3C 273 was the only
quasar detected in $\gamma$--rays (by the {\it COS--B} satellite, \cite{bignami81}). 
When EGRET discovered that blazars were strong $\gamma$--ray emitters as a class \cite{hartman99},
it was also discovered that the optical, the X--ray and the $\gamma$--ray fluxes often varied
simultaneously, shifting the ``inhomogeneous jet" paradigm to the
simpler ``one--zone" model.
Having fewer parameters, this model played a crucial role for our understanding of jets,
and this is the reason of its success.

However, we should not forget that this model adopts a drastic simplification: after all we do see,
also in the radio VLBI maps, that the jet is composed of several distinct knots,
and it is therefore very likely that also on the sub--pc scale there is
more than one zone contributing to the observed flux, even if a single one 
may dominate in one particular frequency range.
These sub--components could vary independently, and possibly have a lifetime
comparable to the light crossing time, since the cooling time (both radiative
and adiabatic) of their radiating particles is short.
At any given time, we may see one or more than one component.
Since the locations of these components will be different, 
their size will also be  different, as well as their magnetic field
and the amount of the external radiation they see.

\begin{figure}
\hskip -0.3 cm
\includegraphics[width=8.8cm,clip]{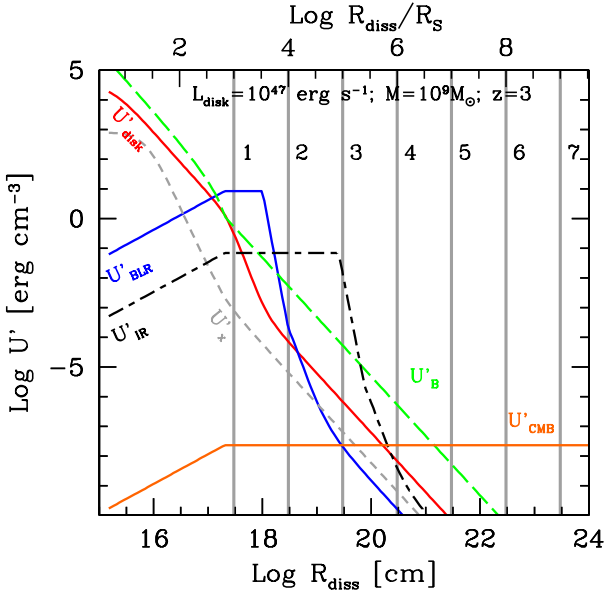}\\
\vskip -1.1 cm
\caption{
Energy densities as 
seen in the comoving frame of the emitting blob as a function of 
the distance from the black hole (in \sch\ radius units in the upper x--axis).
We assumed: $M=10^9 M_\odot$, $L_{\rm d}=10^{47}$ erg s$^{-1}$, $z=3$ 
and $\Gamma =\min[15, (R_{\rm diss}/3R_{\rm S})^{1/2}]$.
$U^\prime_B$: magnetic energy density;
$U^\prime_{\rm d}$: radiation coming directly from the disk,
assumed to be a standard Shakura--Sunyaev disk \cite{ss73};
$U^\prime_{\rm BLR}$ and $U^\prime_{\rm IR}$: contribution from the BLR
and the molecular torus, respectively;
$U^\prime_{X}$: contribution from the X--ray corona emission;
$U^\prime_{\rm CMB}$: contributions from the Cosmic Microwave Background.
The grey vertical lines indicates the distances used to construct
the SEDs shown in Fig. \ref{sedrdiss}.
}
\label{rdiss}       
\end{figure}
\begin{figure}
\hskip -0.3 cm
\includegraphics[width=8.8cm,clip]{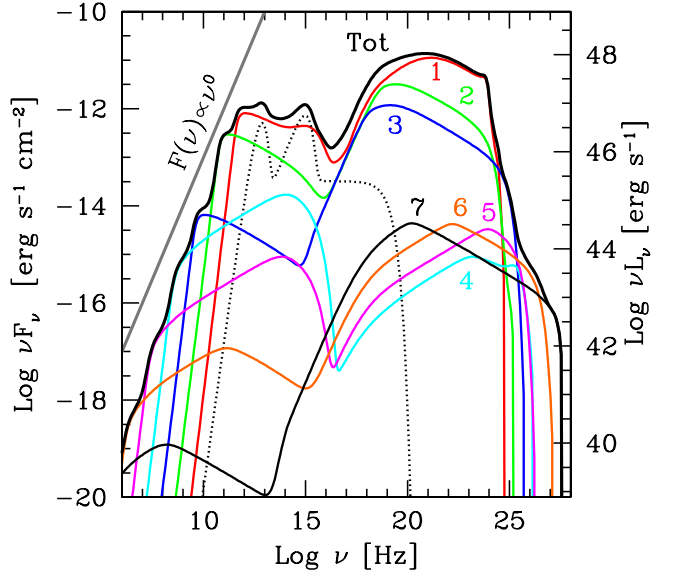}
\vskip -0.7 cm
\caption{
Sequence of SEDs calculated for different $R_{\rm diss}$ 
from $10^3 R_{\rm S}$ to $10^9 R_{\rm S}$ (one per decade).
The injected electron luminosity is $P^\prime_{\rm i}=10^{44}$ erg s$^{-1}$
for $R_{\rm diss}=10^3 R_{\rm S}$ and is reduced by a factor 3 each decade.
The particle distribution has always the same $\gamma_{\rm b}=100$, while 
$\gamma_{\rm max}$ increases by a factor 3 each decade starting
from $\gamma_{\rm max}=10^4 $ for $R_{\rm diss}=10^3 R_S$.
The shown spectra neglect  the absorption of the 
high energy flux due to the IR--opt--UV cosmic background.
The numbers correspond to the same numbers of Fig. \ref{rdiss}.
The thicker black line is the sum of all the SEDs.
The grey line at radio frequencies indicates $F(\nu) \propto \nu^0$.
The flux is calculated assuming $z=3$. Adapted from \cite{gg09d}.
}
\label{sedrdiss}       
\end{figure}

\subsection{Location of the main emission region}

\begin{figure}
\vskip -0.2 cm
\hskip -0.4 cm
\includegraphics[width=9cm,clip]{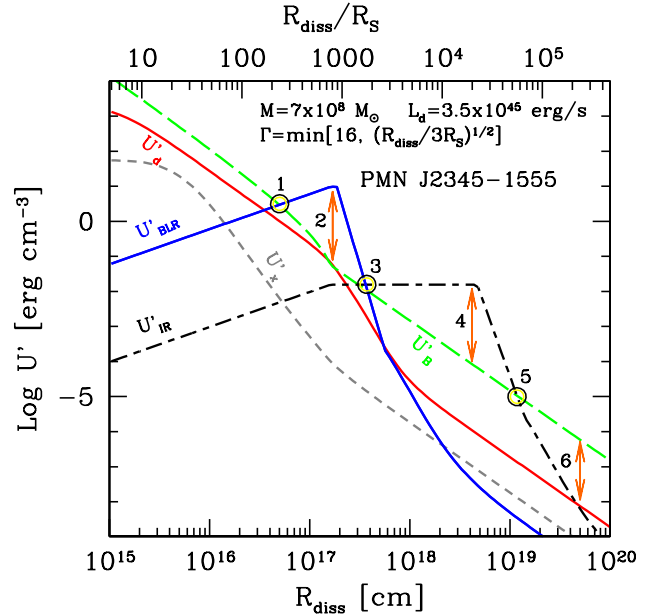}
\vskip -0.5 cm
\caption{Similar to Fig. \ref{rdiss}, to illustrate the distances where
the largest and the smallest Compton dominance are predicted.
The used parameters are for PMN J2345--1555, the FSRQs that became ``blue" 
during a flare \cite{gg13a}.
The numbers next to the circles and arrows correspond to the location where
$U^\prime_{\rm B}$ is equal to the dominant form of radiation energy density
(points 1, 3 and 5), and where instead the ratio between the radiative and 
magnetic energy density is maximized (arrows 2, 4)  leading to the 
maximum expected Compton dominance.
Arrow 6 corresponds instead to a synchrotron dominated SED.
}
\label{u2345}       
\end{figure}

In \cite{gg09d}, we studied how the different radiation energy
densities change as a function of the distance $R_{\rm diss}$ from
the black hole. 
This is illustrated in Fig. \ref{rdiss}.
This figure shows the energy densities as observed in the comoving frame of
a jet that is accelerating (as $\Gamma\propto R^{1/2}$, see e.g. \cite{vlahakis04}) 
in its inner part,
until it acquires a final value of $\Gamma$.
We also assume that the broad line region (BLR) re--emits 10\% of the 
luminosity produced by the disk in the form of broad emission lines.
The BLR is assumed to be a spherical shell of radius 
$R_{\rm BLR} \propto L_{\rm d}^{1/2}$.
Since $U^\prime_{\rm BLR} \sim L_{\rm BLR}\Gamma^2 /[4\pi R_{\rm BLR}^2 c]$
we have that within the BLR $U^\prime_{\rm BLR} \sim \Gamma^2/[12\pi]$,
i.e. {\it it depends only on $\Gamma$}.
Beyond the BLR the line seed photons are seen from behind, and
are more and more de--beamed as the distance increases.
This makes $U^\prime_{\rm BLR}$ to decrease faster than $R^{-2}$.
In analogy with radio--quiet quasars, we calculate the contribution
to the external radiation energy density of a molecular torus, intercepting
a substantial fraction of the disk luminosity and re--emitting it in the IR \cite{sikora02}.
The distance from the black hole of this structure is much larger than $R_{\rm BLR}$
and the corresponding $U^\prime_{\rm IR}$ is smaller.
For the magnetic field, we simply assume that the Poynting flux 
$L_{\rm B}\propto R^2_{\rm diss} \Gamma^2 B^2$ 
carried by the jet is a constant fraction of the jet total power.
For a constant $\Gamma$ this implies $B\propto R^{-1}_{\rm diss}$.  
At large distances (i.e. several kpc away) the dominant contribution to 
the external radiation energy density becomes the Cosmic Microwave Background (CMB),
with a corresponding $U^\prime_{\rm CMB} \propto \Gamma^2 (1+z)^4$.  
These simple and reasonable scalings play a major role in shaping the SED.
In fact, for different $R_{\rm diss}$, i.e. the location where the jet transforms part
of its power into radiation, the ratio 
$U^\prime_{\rm rad}/U^\prime_{\rm B}\approx L_{\rm C}/L_{\rm syn}$
is very different. 
The Compton dominance, defined as the ratio between the inverse Compton to 
synchrotron luminosities, $L_{\rm C}/L_{\rm syn}$, is maximized  
at specific distances.
To better illustrate this point, we have labelled in Fig. \ref{u2345}
the distances $R_{\rm diss}$ of maximum Compton dominance (labelled 2 and 4).
We expect $L_{\rm C}/L_{\rm syn}\sim 1$ at the distances labelled 1, 3 and 5.
A synchrotron dominated SED is expected only beyond point 5, with a 
minimum Compton dominance at distance 6. 
Beyond that, if the jet continues to remain relativistic with the same $\Gamma$, 
$U^\prime_{\rm CMB}$ takes over and  $L_{\rm C}/L_{\rm syn}\gg 1$ again.
Going back to Fig. \ref{rdiss}, we can calculate the SED resulting from dissipation at 
different $R_{\rm diss}$, with the following recipe:
\begin{enumerate}

\item we calculate the SED each decade of $R_{\rm diss}$,
from $R_{\rm diss}=10^3R_{\rm S}$ to $R_{\rm diss}=10^9R_{\rm S}$, as labelled in Fig. \ref{rdiss};

\item the power $P^\prime$ in relativistic electrons injected into 
the source scales as $R_{\rm diss}^{-1/2}$;

\item relativistic electrons are injected according to a 
power law $Q(\gamma)\propto \gamma^{-2.5}$ between a constant $\gamma_{\rm min}=100$ and
$\gamma_{\rm max}\propto R_{\rm diss}^{1/2}$, starting from $\gamma_{\rm max}=10^4$ at 
$R_{\rm diss}=10^3 R_{\rm S}$;

\item the Poynting flux remains constant after the acceleration region.
This implies $B^2\Gamma^2R^2_{\rm diss}=$const;

\item the particle distribution is calculated through the continuity equation at a
time equal to the light crossing time which is also the dynamical time, namely the time
required to double the size of the source. 
Continuos injection of relativistic particles, radiative cooling and electron--positron pair production
are accounted for.
\end{enumerate}
With the above prescriptions, the density $n$ of the relativistic particles 
scales approximately as $n\propto R^{-2}_{\rm diss}$, while $B\propto R_{\rm diss}^{-1}$.
If the jet were continuous, and not fragmented into several blobs,
we would obtain a {\it flat radio spectrum}, with a self--absorption frequency
scaling as $R^{-1}_{\rm diss}$. 
The sum of the emission of our discrete components (black thick line in Fig. \ref{sedrdiss}) 
is not far from this, as can be appreciated comparing with the
grey oblique line indicating a $F(\nu)\propto \nu^0$ spectrum.
The SEDs in Fig. \ref{sedrdiss} show a great
diversity and a non--monotonic behavior, but this is in agreement with the 
$U^\prime_{\rm rad}/U^\prime_{\rm B}$ ratio shown in Fig. \ref{rdiss}.
Particularly interesting is the fact that $\gamma$--ray spectrum in the
GeV band is dominated by the innermost emitting zone, while the
X--ray flux receives the contributions from the first 3 zones.
This would imply a variability of the GeV flux more pronounced and faster  
than the X--ray flux, in agreement with what observed in powerful blazars
where the [0.3--10 keV] flux is produced by the inverse Compton process.
In these examples, the synchrotron self--Compton process is unimportant
for all SEDs but the number 4, which is the one with the smallest
Compton dominance. 
At larger distances, in fact, $U^\prime_{\rm CMB}$ becomes larger than $U^\prime_{\rm B}$ 
and the Compton dominance increases again. 
These large regions of the jet emit most of their flux in the MeV--TeV band.

To conclude this section, note that the most efficient location to produce
the largest amount of $\gamma$---rays is at $\sim 10^3$ and at $10^4$ \sch\ radii,
corresponding to 0.1 and to 1 pc.
In these two locations there is the largest amount of seed photons and the maximum $\Gamma$.
At smaller radii the disk radiation may be even larger, but the beaming is modest,
at larger radii the broad lines or the IR photons coming from the torus 
are seen coming from the back, and are not boosted any longer.

This apparently reasonable conclusion is challenged by two facts.
The first is that in some sources there is a correlation between the 
radio flares, the switch in polarization angle and the occurrence of
$\gamma$--ray flares (see, e.g. \cite{jorstadt13}, \cite{marscher08}),
suggesting that also the $\gamma$--ray emission originates in a region
more than one parsec away from the black hole.
The second issue is the variability of the TeV emission,
which in some source is embarrassingly fast, as discussed below.

\subsection{Fast TeV variability, a puzzle}

Ground based Cherenkov telescopes collect enough photons to
study fast variability of bright blazars, even on timescales of minutes.
And they did discover that the TeV flux is sometimes varying
on these timescales, which is the shortest observed at any wavelength.
The BL Lac objects
Mkn 501 ($z=0.0336$) and PKS 2155--304 ($z=0.116$) 
showed variations on $t_{\rm var}=$3--5 minutes of their TeV flux
(\cite{albert07}, \cite{aharonian07}).
This challenged models in which the minimum variability
timescales is related to the light crossing time of the \sch\ radius
of the black hole of the system (see e.g. \cite{begelman08}),
and strongly suggested that the region emitting these ultrafast
TeV flares is very compact, smaller than the cross section radius of the jet 
(\cite{giannios09}, \cite{gg09c}, \cite{marscher10}, see also \cite{lyutikov06} for a
similar problem in Gamma--Ray Bursts). 
Nevertheless the location of
this TeV emitting region could still be at $R_{\rm diss}\sim 10^3 R_{\rm S}$,
where we think that most of the dissipation is taking place.
In this case we have a ``jet in a jet", or rather ``needle in a jet"
scenario, where the compact TeV emitting zone is immersed in a larger
and active jet zone, and can use the seed synchrotron 
photons produced by the larger region to produce the TeV flux.
This is allowed by the absence, in these BL Lac objects, of broad emission lines:
while helping the very production of MeV--GeV photons 
(because they are used as seeds for Compton scatterings), the line photons 
act as killers of $\gamma$--rays above a few tens of GeV, through photon--photon
collisions (see e.g. \cite{liu06}, \cite{tavecchiomazin09} and \cite{poutanen10})

According to the above idea, FSRQs (with prominent emission lines) {\it should not}
be strong TeV emitters, and instead a few of them are.
Up to now we know 3 TeV FSRQs:
3C 279 ($z=0.536$) \cite{albert08}; \cite{aleksic11a};
PKS 1510--089 ($z=0.36$) \cite{wagner10}, and
PKS 1222+216 ($z=0.431$) \cite{aleksic11b}.
To these sources one can add PKS B1424--418 ($z=1.52$), not detected in the TeV band, but
with photons above tens of GeV detected by {\it Fermi} \cite{tavecchio13} varying on a
$\sim$day timescale.

PKS 1222+216 doubled its TeV flux in just 10 minutes, constraining
the emitting region size to be $R<c t_{\rm var}\, \delta/(1+z) 
\simeq 5\times 10^{14} (\delta/20)$ cm 
($\delta$ is the relativistic Doppler factor). 
This is an extremely compact region, and yet cannot be located within the BLR,
whose photons would absorb the emission above $\sim$20 GeV.
The TeV emitting region must be located outside.
This remains true even if the BLR has a flattened geometry (as suggested by e.g. \cite{shields78};
\cite{jarvis06} and \cite{decarli11}), as shown in \cite{tavecchioinprep}. 

At $R>R_{\rm BLR}$ the radiation field is dominated by the thermal 
radiation of the dusty torus \cite{blazejowski00} and the opacity is smaller.
We must conclude that the varying TeV emitting region is
both small and very distant from the black hole.
It is a bullet (or a bomb) with a milli--pc size at a distance of more than a pc.
Ideas to explain it involve  
strong recollimation and focusing of the flow (e.g. \cite{stawarz06}, 
\cite{bromberg09}, \cite{nalewajko09} or complex reconnection events \cite{giannios13},
or the existence of new particles, like axions: in this case high energy photons
could be produced within the BLR, be converted into axions, insensitive to $\gamma$--$\gamma\to$ e$\pm$
processes, and converted again into photons at large distances \cite{tavecchio12}.

\subsection{Tentative summary}

Most of the time and in most of the sources, 
there is one preferred location in the jet where most of the jet bolometric luminosity
is produced, from the IR to the $\gamma$--rays.
Usually this is within the BLR, at $\sim 10^3 R_{\rm S}$, rarely beyond.
This can be proved by the presence of a break at $\sim$10 GeV due to $\gamma$--$\gamma \to$ e+e--
\cite{poutanen10} in bright $\gamma$--ray FSRQs with enough photons above 10 GeV.
Other emitting regions are however likely.  
Even if their bolometric output is less, they may significantly contribute to
the flux in specific bands, especially in the soft X--rays and the optical.

Sometimes much smaller jet regions become active, 
not only at a distance $\sim 10^3 R_{\rm S}$ from the black hole,
but also much further out.
The ratio between the distance and their size can approach $10^3$
(i.e. one parsec over a milli--parsec).
The geometry and the emission process of these regions is an open issue.
However we can infer that their spectrum is peaked at high $\gamma$--ray energies,
since we never saw extremely fast variability at other well--sampled
frequencies, such as soft X--rays.

The presence of sub--jet regions is also inferred by the X--ray emission
observed by the {\it Chandra} satellite at hundreds of kpc away from the nucleus
(see e.g. \cite{schwartz10} for review).
This emission is interpreted as inverse Compton scattering of relativistic electrons off
the CMB photons, requiring the jet to have a bulk Lorentz factor of $\sim$10 also
at that distances.
The radiative cooling time of the emitting electrons is very long, and 
also the adiabatic cooling timescale is long if one assumes that the emission
comes from a blob with a radius equal to the entire cross section of the jet.
If so, the X--ray knot should be resolved by {\it Chandra}, and instead it is not.
A solution to this puzzle is that the emitting size is much smaller than the
cross sectional radius of the jet, to let the emitting electrons loose energy
by adiabatic losses before they cross the jet \cite{tavecchio03}.



\section{Black hole masses of blazars}

\subsection{Virial methods}

Virial methods have become popular to measure the black hole mass.
They are based on the assumption that gravity controls the velocity of the clouds of the BLR.
Through reverberation mapping (\cite{blandford82}; \cite{peterson93}) one estimates $R_{\rm BLR}$, while 
the Full Width Half Maximum (FWHM) of the broad lines is associated to the velocity
$v_{\rm BLR}$ of the clouds ($v_{\rm BLR}= f\cdot$ FWHM, with $f$ of order unity).
The black hole mass is found through $M = R_{\rm BLR} v^2_{\rm BLR}/G$.
The number of AGN with a BLR size measured directly through reverberation mapping
is very small, but good correlations exist between $R_{\rm BLR}$ and
the ionizing luminosity of the accretion disk (\cite{kaspi07}; \cite{bentz13}),
or to the monochromatic disk luminosity at specific frequencies
(with the implicit assumption that the disk spectrum does not greatly change
from one object to the other).
A single spectrum then allows to measure both the FWHM of the line and  
the ionizing disk luminosity, and then the black hole mass.
The uncertainty of these estimates is large, of the order of 0.5--0.6 dex
\cite{vestergard06}, \cite{park12}.
In addition, there may be systematic uncertainties related to the BLR 
geometry (\cite{decarli08}, \cite{decarli11}) 
and the role of radiation pressure \cite{marconi08}.
This large uncertainty is one of the reasons why there is still a debate
about the difference in black hole masses in radio--loud and radio--quiet sources.
A correlation between black hole mass and radio--luminosity
was claimed by \cite{franceschini98} and \cite{laor00}, but
instead \cite{woo02} found that both
the radio--quiet and radio--loud AGN span the same range of black
hole mass, with no evidence for a correlation between radio loudness and black hole mass.
However, \cite{chiaberge11}, taking into account radiation pressure to compute 
the virial black hole masses, recently found that $M>10^8 M_\odot$
is required to produce a radio--loud AGN.

\subsection{Accretion disk fitting}

This was the first method (in the 80s) to measure the black hole mass and the accretion
rate in quasars \cite{shields78}, \cite{malkan83}, but was not widely used thereafter
(see {\cite{calderone13} for a critical discussion).
A revival of this method occurred recently, mainly because
i) the availability of large sample of distant quasars: often this allows
to observe directly the redshifted peak of the accretion disk emission in the optical;
ii) the better knowledge of the different components contributing
to the IR--UV spectrum of a typical quasar;
iii) the realization that even when the peak emission is not seen,
the broad lines can give an estimate of the total disk luminosity.
The latter point is particularly useful for blazars, where the accretion
disk continuum can receive a substantial contribution from the
non--thermal jet component.

The simplest disk model to fit the accretion disk spectrum is
the Shakura \& Sunjaev \cite{ss73} model.
It assumes a \sch\ black hole, a geometrically thin, optically thick disk 
that emits as a blackbody at each radius.
Although these assumptions appear rather crude at first sight, they are
not as bad as they seem \cite{calderone13}, and can give a good estimate
of the black hole mass of the object.
The uncertainty associated to this method depends on the data quality:
if the peak of the disk emission is visible, then the uncertainty on the
black hole mass is less than a factor 2.

\subsection{Blazars and accretion disk emission}

Powerful blazars are ``red", with a synchrotron component
peaking in the sub--mm band and steep in the IR--optical.
This leaves their accretion disk component ``naked", and
well visible.
To illustrate this point, Fig. \ref{0836} shows the SED of PKS 0836+710 ($z=2.17$).
The labels indicate the different components: the dashed black line
correspond to the accretion disk, the IR torus and the X--ray corona emissions;
the green solid line is the synchrotron component, the long dashed
grey line is the SSC contribution, and the dot--dashed grey line is the EC 
component, dominating the high energy hump.
The disk emission component, and its peak, are clearly visible.
In these cases it is possible to derive $M\sim 3\times10^9M_\odot$,
and $L_{\rm d}\sim 2\times10^{47}$ erg s$^{-1}$, nearly half
the Eddington luminosity \cite{gg10a}.
This source is present in the catalog of {\it Fermi} sources
that collects sources detected during the first 2 years of the mission, \cite{ackermann11},
but not in the 1LAC (first year, \cite{abdo10}), not in the LBAS (first 3 months, \cite{abdo09}),
even if it is one of the most powerful blazar.
On the contrary, it has been detected during the 3--years survey of the BAT instrument
onboard {\it Swift} \cite{ajello09}.

Fig. \ref{1023} shows the SED of the second most distant blazar known, B2 1023+25,
at $z=5.3$,
selected as a blazar candidate because of its radio brightness and large radio--loudness,
and soon after confirmed as a blazar by {\it Swift} X--ray observation in the 0.3--10 keV band
\cite{sbarrato12b}.
Although the received X--ray photons were a few, they suggest a large flux and a hard
spectrum, two signatures of a blazar.
B2 1023+25 was selected within the $\sim8,800$ square degrees covered both by the SDSS (Sloan Digital Sky Survey)
and FIRST (Radio Images of the Sky at Twenty-Centimeters). 
It implies the existence of other $\sim$3 similar blazars (all sky) and then $\sim$1800 intrinsically similar 
blazars, but misaligned (all sky).
The fact that its black hole mass is $M\sim 3\times 10^9M_\odot$ implies the existence
of other $\sim$1800 black holes in jetted sources as heavy as that at the same redshift.
At $z\sim$5.3, the Universe was only one billion years old.
Through these numbers one can understand why the search for heavy black holes in blazars
can help the study of black formation and growth at high redshift, even challenging similar studies
based on searches of radio--quiet objects.

\begin{figure}
\vskip -0.5 cm
\hskip -0.3 cm
\includegraphics[width=8.8cm,clip]{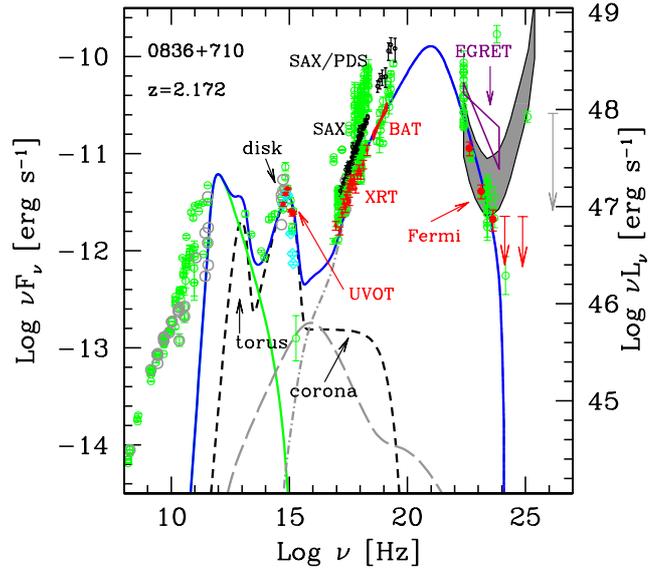}
\vskip -0.5 cm
\caption{
The SED of the blazar PKS 0836+710, together with the fitting model,
to show the different contributions of the accretion disk, the molecular torus and the
X--ray corona (short dashed line); the synchrotron component (green solid line); 
the synchrotron Self--Compton
component (long dashed line) and the External Compton component (dot--dashed line).
This is the typical SED of a very powerful FSRQ. 
It can be seen that the synchrotron spectrum peaks in the sub--mm band and is steep above the peak,
as steep as the $\gamma$--ray emission (because the model assumes that the fluxes in both bands
are produced by the same electrons).
{\it This leaves the emission disk component ``naked" and visible,} opening the possibility 
to measure the mass and the accretion rate of the black hole.
Adapted from \cite{gg10a}.
}
\label{0836}       
\end{figure}

\begin{figure}
\vskip -0.5 cm
\hskip -0.3 cm
\includegraphics[width=8.8cm,clip]{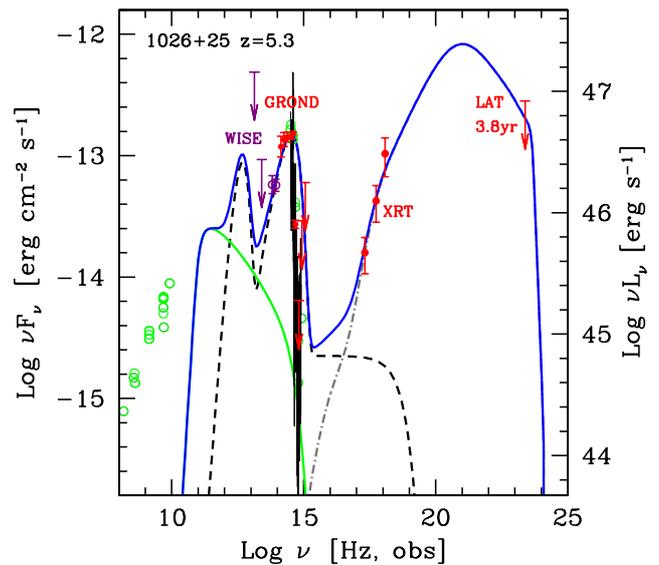}
\vskip -0.5 cm
\caption{
The SED of B2 1023+25, the second most distant blazar at $z=5.3$,
together with a fitting model.
Note that the WISE, GROND and SDSS data can be fitted by a Shakura \& Sunjaev \cite{ss73}
accretion disk, yielding a black hole mass $M=2.8\times 10^9 M_\odot$ and a disk luminosity
$L_{\rm d}= 9\times 10^{46}$ erg s$^{-1}$, equivalent to $L_{\rm d}/L_{\rm Edd}=0.25$. 
From \cite{sbarrato12b}.
}
\label{1023}       
\end{figure}

\subsection{Early and heavy black holes}

According to the blazar sequence, and to the evidences gathered
so far, powerful blazars emit most of their luminosity
in the high energy hump, which peaks between 1 and 10 MeV.
Being so powerful, these blazars can be detected also at high redshifts,
and thus shed light on the far Universe.
The two main instruments best suited to find 
blazars at high redshifts are BAT (15--150 keV), onboard {\it Swift} and 
LAT (0.1--100 GeV), onboard {\it Fermi}.
The {\it INTEGRAL} satellite is somewhat more sensitive than BAT in the
same energy range, but a significant fraction of the exposure time is dedicated
to observations of sources in the Galactic plane, while BAT, which follows Gamma Ray Bursts,
is observing more uniformly the entire sky.

BAT, after the first 3 years of operations, detected 38 blazars.
10 of them have $z>2$.  Of these, 5 have $z>3$.
LAT instead detected 31 blazars at $z>2$ in two years, but only 2 have $z>3$.
These numbers, albeit small, already suggest that the hard X--ray band
is more efficient than the GeV band to find out high--redshift powerful blazars.
In \cite{gg10a} and \cite{volonteri11} we have considered all 10 BAT blazars at $z>2$, 
finding that all of them have black hole masses $M>10^9M_\odot$,
and all of them have a [15--55 keV] luminosity greater than $1.5\times 10^{47}$ erg s$^{-1}$.
The latter point allowed to associate the luminosity function of BAT blazars \cite{ajello09} to
the mass function of heavy (i.e. $M>10^9M_\odot$) and active ($L_{\rm d}>0.1 L_{\rm Edd}$)
black holes.
Conservatively, we have also modified the luminosity function in \cite{ajello09} adding
an exponential cut above $z=4$, where \cite{ajello09} had no data.
In the 4--5 redshift bin, however, we have 4 serendipitously discovered blazars
with $M>10^9M_\odot$ \cite{yuan05}, \cite{hook95}, \cite{worsley04a} and \cite{worsley04b},
and two blazars at $z>5$: one is the most distant blazar known, Q0906+6930 at $z$=5.47,  \cite{romani06}
(green pentagon in Fig. \ref{fim}) and the other is the recently discovered
blazar B2 1023+25 \cite{sbarrato12b}, which was discovered through a selection within the SDSS+FIRST survey
(yellow pentagon).

For each detected blazar, there must be other $2\Gamma^2=450(\Gamma/15)^2$ 
other sources pointing away from us, 
and we have adopted $\Gamma=15$ as the typical bulk Lorentz factor of the relativistic jets
\cite{gg10b}.
The result showed that the mass function of heavy black holes, as a function of redshift,
had a peak at $z\sim 4$ (see the red stripe in Fig. \ref{fim}).
This was surprising, since the mass function of heavy and active 
black holes in radio--quiet objects has a peak at $z\sim 2$  
(blue lines in Fig. \ref{fim}).
However, for the jetted sources, we had considered only BAT blazars.
There could be the possibility that many more heavy and active black holes
could be revealed by LAT, possibly preferentially at $z\sim 2$, to re--establish
a symmetry with radio--quiet objects.
But when the $\gamma$--ray luminosity function was derived by \cite{ajello12},
we could calculate the mass function of heavy and active black holes of LAT blazars,
which is illustrated in Fig. \ref{fim} by the green stripe \cite{gg13b}.
There are too few LAT blazars with heavy and active black holes to shift
the peak from $z\sim4$ to $z\sim2$.
We must therefore take this result seriously.
We are led to conclude that most massive black holes forms at two epochs:
in jetted sources they form earlier ($z\sim4$) than in radio--quiet objects ($z\sim 2$).

As a consequence of the two different formation epochs, the ratio of radio--loud to quiet
sources with a heavy black hole is strongly evolving.
We stress that this refers only to those sources hosting a black hole with $M>10^9M_\odot$,
not to the entire population of quasars.

\begin{figure}
\vskip -0.5 cm
\hskip -0.3 cm
\includegraphics[width=9cm,clip]{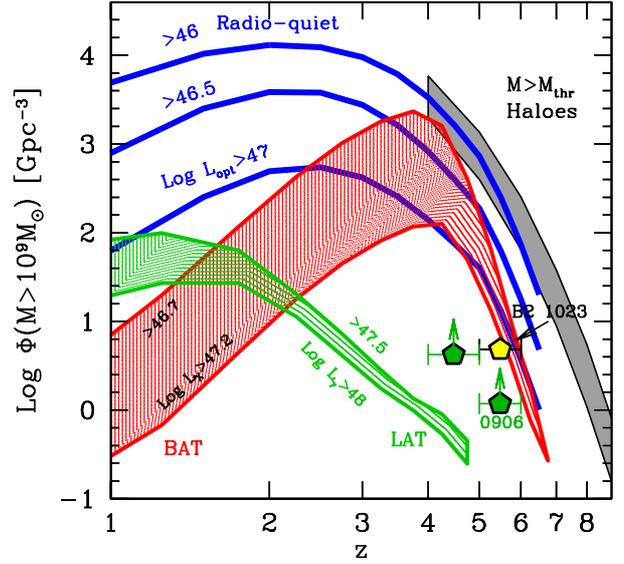}
\vskip -0.8 cm
\caption{
Comoving number density of  blazars powered by ``heavy and active" black holes 
($M>10^9M_\odot$, $L_{\rm d}/L_{\rm Edd}>0.1$) as a function of redshift.
The larger red hatched band is derived by integrating the [15--55 keV]
luminosity function \cite{ajello09} 
modified as in \cite{gg10a} 
above  $\log L_X =46.7$ (upper boundary) and
$\log L_X =47.2$ (lower boundary), and multiplying
the derived density by 450 (i.e. $2\Gamma^2$, with  $\Gamma=15$).
The smaller green hatched band is derived by integrating the 
$\gamma$--ray luminosity function above 
$\log L_\gamma=47.5$ (upper boundary) and
$\log L_\gamma =48$ (lower boundary). 
All sources selected should own a disk with $L_{\rm d}>0.1L_{\rm Edd}$, 
accreting onto a SMBH of mass $>10^9 M_\odot$.  
The three (blue) stripes are derived integrating the luminosity function of
radio--quiet quasars \cite{hopkins07}, above three different threshold
luminosities, as labelled (see also \cite{willott00} and \cite{volonteri11}).
The grey stripe is based on connecting black hole mass to halo 
mass, as described in \S 7.2 of \cite{gg10a}; $M_{\rm thr}$
is the minimum halo mass required to host a $10^9 M_\odot$ black hole. 
This can be considered as the upper limit to the density of $10^9 M_\odot$ black holes.
The (green) pentagons correspond to the density inferred from the few 
sources at high--$z$ already identified as blazars.
The (yellow) pentagon labelled B2 1023 is the density inferred from the existence of only one blazar,
B2 1023+25, in the region of the sky covered by the SDSS+FIRST surveys \cite{sbarrato12b}.
Adapted from \cite{gg13b}.
}
\label{fim}       
\end{figure}

\begin{figure}
\vskip -0.5 cm
\hskip -0.3 cm
\includegraphics[width=9cm,clip]{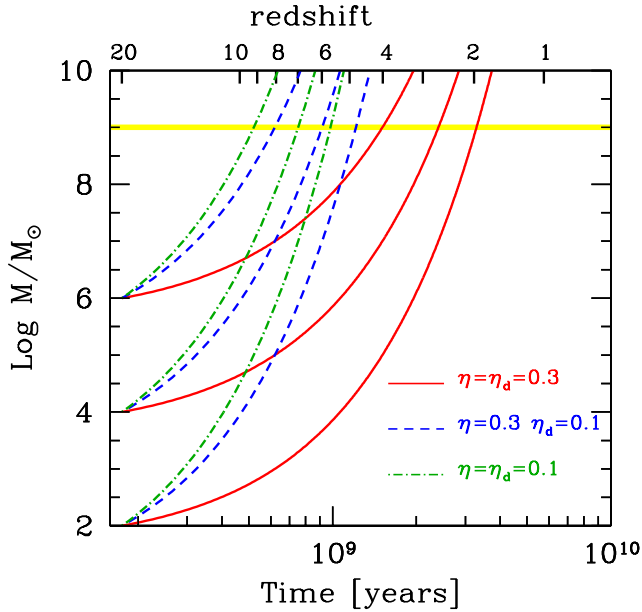}
\vskip -0.5 cm
\caption{
The mass of a black hole accreting at the Eddington rate as
a function of time (bottom axis) and redshift (top axis). 
Accretion starts at $z=20$ onto a black hole seed of $10^2M_\odot$,
 $10^4M_\odot$ or $10^6M_\odot$ , with different efficiencies, as labeled.
The horizontal line marks one billion solar masses.
The larger $\eta_{\rm d}$, the smaller the amount of accreted mass
needed to produce a given luminosity, and the longer the black hole 
growing time. If part of the accretion energy goes into launching a jet,
however, $\eta_{\rm d}<\eta$ and the growth time decreases. From \cite{gg13b}.
}
\label{salpeter}       
\end{figure}

\subsection{Black hole growth and spin}

It is popular to associate the presence of a relativistic jet with a
black hole spinning rapidly.
In fact, the rotational energy of the hole can be an important 
source of power for the jet (\cite{blandford77}, \cite{tchekhovskoy11}).
This led to the idea that the radio--loud/radio--quiet dichotomy is associated 
to the value of the spin (\cite{wilson95}, \cite{sikora07}).
Since for spinning black holes the innermost stable orbit moves inward as the
spin increases, the accretion efficiency also increases, to reach a theoretical maximum
of $\eta=0.42$ for a dimensionless spin value $a=1$ (here $\eta$ is
defined by $L_{\rm d} =\eta \dot M c^2$, where $\dot M$ is the mass accretion rate).
However, \cite{thorne74} pointed out that if the disk is emitting, there is
a torque exerted by the photons on the hole, inhibiting it to spin with $a=1$.
The maximum value is $a=0.998$, corresponding to a maximum $\eta=0.3$.
This means that, with respect to a \sch\ black hole with $\eta \sim 0.06$, 
a maximally spinning hole allows to reach the Eddington luminosity with 
1/5 of the accretion rate.
As a consequence, if accretion is Eddington limited, the spinning black hole will grow at
a slower rate.
This simple argument leads to a crucial problem:
if jets are associated to large spins, their black holes should grow at a slower rate than
the hole in radio--quiet quasars.
At high redshifts, there should be no jetted sources with a large black hole mass.
Quite the opposite of what observed.

Fig. \ref{salpeter} illustrates the point.
It assumes that the black hole starts to accrete at $z=20$ with a seed mass 
of $10^2$, $10^4$ or $10^6 M_\odot$, and accretes at the Eddington rate with 
an accretion efficiency $\eta_{\rm d}=0.3$ (red solid lines)
or $\eta_{\rm d}=0.1$ (green dot--dashed lines).
If $\eta=0.3$, the black hole mass reaches $10^9 M_\odot$ at $z=2$ for a $10^2 M_\odot$ seed,
and at $z=4$ for a $10^6M_\odot$ seed. 
Even a very large black hole seed is not enough to account for what we see: a black hole
mass exceeding $10^9M_\odot$ in a jetted source at $z=5.3$.
We conclude that one of the assumptions usually made to calculate the black 
hole growth must be wrong.

In \cite{gg13b} we suggested a possible way out of this problem, by
considering the possibility that the gravitational energy of the accreted matter
does not entirely heat up the disk, but it is partly used to amplify the magnetic
field that can act as a catalyzer for the jet formation and acceleration.
In other words, for a given accretion rate, the produced disk luminosity is smaller,
because it corresponds to a fraction of the total efficiency.
Following \cite{jolley08} (see also \cite{shankar08}), 
the global efficiency $\eta$ is the sum
of two terms, one for the generated magnetic field ($\eta_{\rm B}$), 
and the other for the disk luminosity ($\eta_{\rm d}$):
\begin{equation}
\eta \, \equiv \, \eta_{\rm B} +\eta_{\rm d}.
\end{equation}
A rapidly spinning hole can have $\eta=0.3$, but the fraction of
the gravitational energy going to heat the disk could correspond to
$\eta_{\rm d}<0.3$.  
As a consequence, the black hole grows faster, 
doubling its mass on a  Salpeter time \cite{salpeter64}, which is
now modified in \cite{gg13b}:
\begin{equation}
t_{S} \, = \, {M\over \dot M} \, =\,
{\eta_{\rm d} \over 1-\eta } \, {\sigma_{\rm T} c \over 4\pi G m_{\rm p}}
\, {L_{\rm Edd}\over L_{\rm d} }
\label{ts}
\end{equation}
If $\eta_{\rm d}$ becomes small because a fraction of the energy goes
to amplify the magnetic field, then the Salpeter time becomes smaller
and the black hole can grow faster.
This is illustrated in Fig. \ref{salpeter} by the dashed blue lines,
that assume $\eta_{\rm d}=0.1$ and $\eta=0.3$.
In this case  even Kerr holes with maximum spin
and large total efficiency can grow fast enough to reach $M=10^9 M_\odot$ 
at redshifts larger than 5.

The above hypothesis is not implying that accretion is directly the only responsible
for powering the jet. 
Most of the jet power can still come from the rotational energy of the hole.
But this energy, to be extracted efficiently enough, needs large magnetic fields.
These are amplified using part of the gravitational accretion energy.

\section{Summary and conclusions}

\begin{itemize}

\item {\it Blazar sequence so far confirmed by Fermi  ---} 
{\it Fermi} blazars observed up to now, with known redshift, obey the
phenomenological blazar sequence. Studies in the optical--UV of
featureless {\it Fermi} blazars succeeded in setting an upper limit
to their redshift through the absence of Lyman--$\alpha$ absorption.
As a corollary, all blazars observed by {\it Fermi} are at $z<4$.
 
\item {\it Blazar divide ---} 
Blazars can be divided into low  power BL Lacs and high power FSRQs.
This parallels the division in FR I and FR II radio--galaxies.
This division could be the result of a change of the accretion regime:
from radiative inefficient to efficient disks, at a dividing 
disk luminosity approximately 1\% of the Eddington one.

\item {\it One zone vs multizone vs continuous jets ---} 
{\it Most} jets {\it usually} emit most of their luminosity 
in a single region at $\sim 10^3$ \sch\ radii.
It is very likely that there are other emitting regions of the jet,
contributing to a lesser extent in terms of bolometric luminosity,
but that can give important contributions in selected frequency ranges.
In addition, {\it sometimes} there are compact regions of the jet,
much smaller than the cross section radius of the jet itself, that
emit, preferentially at $\gamma$--ray energies, a strong and fast varying flux.
These are required by the fast varying TeV emission in (line emitting) FSRQs.
The emitting region of this radiation should be located beyond the BLR.

\item {\it Accretion disks and blazars ---} 
The most powerful blazars have a ``red" non--thermal SED,
that leaves unhidden the accretion disk emission.
Its luminosity, and frequency peak, give both $M$ and $\dot M$
with an uncertainty often smaller than virial methods.
This allowed to study the disk and jet power as a function 
of the Eddington luminosity.
Relativistic extragalactic jets are present for all 
$L_{\rm d}/L_{\rm Edd}$ values we can sample.
The jet power and the disk luminosity correlate.
This implies that jets and accretion must correlate.
On the other hand, often the jet power is larger than 
the disk luminosity \cite{celotti08}, \cite{gg10b}, \cite{bonnoli11},
implying that accretion {\it cannot be} the only driver of jets.
We believe that accretion amplifies the magnetic field close to the
\sch\ radius. 
This field can then extract the rotational energy of the hole.

\item {\it Heavy and early black holes ---} 
Blazars are powerful and can be seen at large redshifts, so they can 
probe the far Universe.
Since the disk emission is clearly visible in very powerful blazars,
we can measure the black hole mass of these object at large redshifts.
For each source classified as blazars, with a viewing angle less than $1/\Gamma$,
there are other $2\Gamma^2$ similar sources pointing elsewhere.
This makes the hunt for high--$z$ blazars risky (they are a few), but rewarding.

\item {\it Two formation epochs of heavy black holes? ---} 
Powerful blazars emit most of their power at high energies,
in the hard X--rays and in the $\gamma$--ray band.
For the very powerful, the hard X--ray band is closer to their
frequency peak than the [0.1--10 GeV] $\gamma$--ray band.
Hard X--ray surveys are then most productive to find high--$z$ blazars.
Studying their hard X--ray luminosity function, and converting it to 
a black hole mass function, one finds that active ($L_{\rm d}>0.1 L_{\rm Edd}$)
black holes with a mass in excess of a billion solar masses have a peak in their space density
at $z\sim 4$.
This does not change by adding blazars coming from the {\it Fermi} satellite survey.
This contrasts with active and massive black holes in radio-quiet quasars,
whose space density peaks at $z\sim 2$.
We conclude that there are two formation epochs of heavy black holes,
and that jets help the early growth of black holes (or there is a selection effect at action).

\item {\it Black hole growth and spin ---} 
If the presence of a jet is associated to a fast black hole spin, and therefore to 
a large accretion efficiency, we expect that jetted sources accreting at the Eddington limit
need more time to grow that their radio--quiet counterpart.
And yet they are discovered at very large redshift.
This could imply that the global accretion efficiency is in fact the sum
of two parts, corresponding to two way of transforming the available gravitational energy:
one goes to heat the disk, and the other to amplify the magnetic field, that in turn
helps to extract the rotational energy of spinning holes.

\end{itemize}



\vskip 2 cm
\begin{thebibliography}{}
%


\bibitem{abdo09} Abdo A.A. et al., ApJ, {\bf 700}, 597 (2009)   

\bibitem{abdo10} Abdo A.A. et al., ApJ, {\bf 715}, 429 (2010)  

\bibitem{ackermann11} Ackermann M. et al. 
        ApJ, {\bf 743}, 171 (2011) 

\bibitem{albert07} Albert J. et al., 
         ApJ, {\bf 669}, 862 (2007) 
         
\bibitem{albert08} Albert J. et al., Science {\bf 320}, 1752 (2008)  

\bibitem{aharonian07} Aharonian F. et al., 
             ApJ, {\bf 664}, L71 (2007) 

\bibitem{ajello09} Ajello M. et al., 
   ApJ, {\bf 699}, 603  (2009) 
   
\bibitem{ajello12} Ajello M. et al., 
         ApJ, {\bf 751}, 108  (2012) 

\bibitem{aleksic11a} Aleksi{\'c} J. et al., A\&A, {\bf 530}, 4 (2011) 

\bibitem{aleksic11b} Aleksi{\'c} J. et al.,  ApJ, {\bf 730}, L8 (2011) 

\bibitem{begelman08} Begelman M.C., Fabian A.C. \& Rees M.J., MNRAS, {\bf 384}, L19 (2008)

\bibitem{bentz13} Bentz M.C. et al., ApJ, {\bf 767}, 149 (2013)

\bibitem{bignami81} Bignami G.F., et al., 
         A\&A, {\bf 93}, 71, (1981) 

\bibitem{blandford77} Blandford R.D. \& Znajek R.L., MNRAS, {\bf 179}, 433 (1977)

\bibitem{blandford82} Blandford R.D. \& McKee C.F., ApJ, {\bf 255}, 419 (1982) 

\bibitem{blazejowski00} B{\l}a{\.z}ejowski M. et al., 
         ApJ, {\bf 545}, 107 (2000)
         
\bibitem{bonnoli11} Bonnoli G.  et al., 
          MNRAS, {\bf 410}, 368 (2011)

\bibitem{bottcher13} B\"ottcher M. et al., 
          ApJ, {\bf 768}, 54 (2013)   
          
\bibitem{bromberg09} Bromberg O. \& Levinson A., ApJ, {\bf 699}, 1274 (2009)

\bibitem{calderone13} Calderone G. et al., 
         MNRAS, {\bf 431}, 210 (2013)

\bibitem{celotti08} Celotti A. \& Ghisellini G., MNRAS, {\bf 385}, 283 (2008)

\bibitem{chiaberge11} Chiaberge M. \& Marconi A., MNRAS, {\bf 416}, 917 (2011)

\bibitem{decarli08} Decarli R. et al., 
         MNRAS, {\bf 386}, L15 (2008)

\bibitem{decarli11} Decarli R., Dotti M. \& Treves A., MNRAS, {\bf 413}, 39 (2011)

\bibitem{donato01} Donato D. et al.,  
	     A\&A, {\bf 375}, 739 (2001)
	     
\bibitem{fossati98} Fossati G. et al., 
                    MNRAS, {\bf 299}, 433 (1998)

\bibitem{franceschini98} Franceschini A., Vercellone S. \& Fabian A.C., MNRAS, {\bf 297}, 817 (1998)

\bibitem{gg85} Ghisellini G., Maraschi L. \& Treves A., A\&A, {\bf 146}, 204 (1985)

\bibitem{gg98} Ghisellini G. et al., 
         MNRAS, {\bf 301}, 451 (1998) 

\bibitem{gg01} Ghisellini G. \& Celotti A., A\&A, {\bf 379}, L1 (2001)

\bibitem{gg09a} Ghisellini G., Maraschi L. \& Tavecchio F., MNRAS, {\bf 396}, L105 (2009)

\bibitem{gg09b} Ghisellini G. et al., 
         MNRAS, {\bf 399}, L24 (2009) 

\bibitem{gg09c} Ghisellini G. et al., 
         MNRAS, {\bf 393}, L16 (2009) 

\bibitem{gg09d} Ghisellini G. \& Tavecchio F., MNRAS, {\bf 397}, 985 (2009)

\bibitem{gg10a} Ghisellini G. et al., 
         MNRAS, {\bf 405}, 387 (2010)  

\bibitem{gg10b} Ghisellini G. et al., 
         MNRAS, {\bf 402}, 497 (2010)

\bibitem{gg11} Ghisellini G. et al., 
         MNRAS, {\bf 414}, 2674 (2011) 

\bibitem{gg12} Ghisellini G. et al., 
		MNRAS, {\bf 425}, 1371 (2012)
		
\bibitem{gg13a} Ghisellini G. et al., 
	     MNRAS, {\bf 432}, L66 (2013) 

\bibitem{gg13b} Ghisellini G. et al., 
	     MNRAS, {\bf 428}, 1449 (2013) 

\bibitem{giannios09} Giannios D. et al., 
 		MNRAS, {\bf 395}, L29 (2009)

\bibitem{giannios13}	Giannios D.,  MNRAS, {\bf 431}, 355 (2013)

\bibitem{giommi10} Giommi P. et al.   
         MNRAS, {\bf 420}, 2899 (2010)  

\bibitem{giommi12} Giommi P., Padovani P. \& Rau A., 
         MNRAS, {\bf 422}, L48 (2012)  

\bibitem{hartman99} Hartman R.C. et al., 
		ApJS, {\bf 123}, 79 (1999) 

\bibitem{hinton09} Hinton J.A. \& Hofmann W., ARA\&A, {\bf 47}, 523 (2009)

\bibitem{hook95} Hook, I.M. et al., MNRAS {\bf 273}, L63 (1995) 

\bibitem{hopkins07} Hopkins P.F., Richards G.T. \& Hernquist L., ApJ, {\bf 654}, 731 (2007) 

\bibitem{jarvis06} Jarvis M.J. \& McLure R.J., MNRAS, {\bf 369}, 182 (2006)
  
\bibitem{jolley08} Jolley  E.J.D \& Kuncic Z., MNRAS, 386, {\bf 989} (2008) 

\bibitem{jorstadt13} Jorstad S.G. et al., 
         ApJ, {\bf 182}, 147 (2013) 
         
\bibitem{kaspi07} Kaspi S. et al., ApJ, {\bf 659}, 997 (2007)

\bibitem{konigl81} Konigl A.,  ApJ, {\bf 243}, 700 (1981)

\bibitem{laor00} Laor A., ApJ {\bf 543}, L111 (2000)

\bibitem{liu06} Liu H.T. \& Bai J.M., ApJ, {\bf 653}, 1089 (2006)

\bibitem{lyutikov06} Lyutikov M., MNRAS {\bf 369}, L5 (2006)

\bibitem{mahadevan97} Mahadevan R., ApJ, {\bf 447}, 585 (1997)

\bibitem{malkan83} Malkan, M.A., ApJ, {\bf 268}, 582 (1983)

\bibitem{marconi08} Marconi A. et al., 
		ApJ, {\bf 678}, 693 (2008)

\bibitem{marscher80} Marscher A.P., ApJ, {\bf 235}, 386 (1980)

\bibitem{marscher08} Marscher A.P. et al., 
             Nature, {\bf 452}, 966 (2008)

\bibitem{marscher10} Marscher A.P. \& Jorstad S.G. in Proc. {\it Fermi Meets Jansky} (astro--ph/1005.5551) (2010)

\bibitem{morganti92} Morganti R., Ulrich M--H. \& Tadhunter C.N., MNRAS, {\bf 254}, 546 (1992) 

\bibitem{nalewajko09} Nalewajko K. \& Sikora M., MNRAS, {\bf 392}, 1205  (2009)
   
\bibitem{narayan95} Narayan R. \& Yi I., ApJ, {\bf 452}, 710 (1995)

\bibitem{narayan97} Narayan R., Garcia M.R. \& McClintock J.E., ApJ, {\bf 478}, L79 (1997)

\bibitem{nolan12} Nolan P.L. et al., 
 		ApJS, {\bf 199}, 31 (2012)

\bibitem{padovani95} Padovani P. \& Giommi P., ApJ, {\bf 444}, 567, 63 (1995)

\bibitem{padovani07} Padovani P., Ap. and Sp. Science, {\bf 309}, 63 (2007)

\bibitem{park12} Park D. et al., ApJ, {\bf 747}, 30 (2012) 

\bibitem{peterson93} Peterson B.M., PASP, {\bf 105}, 247 (1993)

\bibitem{potter13} Potter W.J. \& Cotter G., MNRAS, {\bf 431}, 1840  (2013)

\bibitem{poutanen10} Poutanen J. \& Stern B., ApJ, {\bf 717}, L118 (2010) 

\bibitem{rau12} Rau A. et al.,  
         A\&A, {\bf 538}, A26 (2012)

\bibitem{romani06} Romani R.W., AJ, {\bf 132}, 1959 (2006) 

\bibitem{salpeter64} Salpeter E.E., ApJ, {\bf 140}, 796 (1964)

\bibitem{sbarrato12a} Sbarrato T. et al., 
         MNRAS, {\bf 421}, 1764 (2012) 

\bibitem{sbarrato12b} Sbarrato T. et al., 
		MNRAS, {\bf 426}, L91 (2012) 

\bibitem{scarpa97} Scarpa R., Falomo R., A\&A, {\bf 325}, 109 (1997)

\bibitem{ss73} Shakura N.I. \& Sunyaev R.A., A\&A, {\bf 24}, 337 (1973)

\bibitem{shankar08} Shankar F. et al., 
         ApJ, {\bf 676}, 131 (2008)

\bibitem{sharma07} Sharma P. et al., 
        ApJ, {\bf 667}, 714 (1007)

\bibitem{shields78} Shields G. A. in Proc. of the {\it Pittsburgh Conf. on BL Lac objects}, Ed. A.M. Wolfe, p. 275 (1978)

\bibitem{shields78} Shields G.A., Nature, {\bf 272}, 706 (1978)

\bibitem{schwartz10} Schwartz D., PNAS, {\bf 107}, 7190 (2010)

\bibitem{sikora02} Sikora M. et al., 
             ApJ, {\bf 577}, 78 (2002)   

\bibitem{sikora07} Sikora M., Stawarz \L. \& Lasota J.--P.,  ApJ, {\bf 658}, 815 (2007)

\bibitem{stawarz06} Stawarz {\L}. et al., 
         MNRAS, {\bf 370}, 981 (2006)

\bibitem{stickel93} Stickel M., Fried J.W. \& K\"uhr H., A\&AS, {\bf 98}, 393 (1993)

\bibitem{tavecchio03} Tavecchio F., Ghisellini G. \& Celotti A., A\&A, {\bf 403}, 83 (2003)

\bibitem{tavecchiomazin09} Tavecchio F. \& Mazin D., MNRAS, {\bf 392}, L40 (2009)

\bibitem{tavecchio10} Tavecchio F. et al. 
        MNRAS, {\bf 405}, L94 (2010) 

\bibitem{tavecchio11} Tavecchio F. et al., 
		A\&A, {\bf 534}, A86 (2011) 

\bibitem{tavecchio12} Tavecchio F, et al., 
		 PhRvD, {\bf 86}, 5036 (2012)	 
	
\bibitem{tavecchio13} Tavecchio F. et al. 
         astro--ph/1306.0734 (2013) 

\bibitem{tavecchioinprep} Tavecchio F. et al., in prep. (2013)

\bibitem{tchekhovskoy11} Tchekhovskoy A., Narayan R. \& McKinney J.C.,  MNRAS, {\bf 418}, L79 (2011)

\bibitem{thorne74} Thorne K.S, ApJ, {\bf 191}, 507 (1974)

\bibitem{urry95} Urry C.M. \& Padovani P., PASP, {\bf 107}, 803 (1995)

\bibitem{vermeulen95} Vermeulen R.C. et al., 
         ApJ, {\bf 452}, L5 (1995)

\bibitem{vestergard06} Vestergaard M.\& Peterson B.M., ApJ, {\bf 641}, 689 (2006)

\bibitem{vlahakis04} Vlahakis N. \& K\"onigl A.,  ApJ, {\bf 605}, 656 (2004) 

\bibitem{volonteri11} Volonteri M. et al., 
         MNRAS, {\bf 416}, 216 (2011)

\bibitem{yuan05} Yuan W. et al., MNRAS, {\bf 358}, 423, (2005) 

\bibitem{wagner10} Wagner S., \& Behera B. in 10th HEAD Meeting, Hawaii (BAAS, 42, 2, 07.05) (2010) 
         
\bibitem{willott00} Willott C.J. et al., 
	      AJ, {\bf 140}, 546 (2000)  

\bibitem{wilson95} Wilson A.S., \& Colbert E.J.M. ApJ, {\bf 438}, 62 (1995)

\bibitem{woo02} Woo J.--H. \& Urry C.M., ApJ, {\bf 581}, L5 (2002)

\bibitem{worsley04a} Worsley M.A. et al., 
         MNRAS, {\bf 350}, 207 (2004a)
		
\bibitem{worsley04b} Worsley M.A. et al., 
         MNRAS, {\bf 350}, L67 (2004b)



\end{thebibliography}
%

\end{document}